\documentclass[fleqn,12pt,twoside]{article}
\usepackage{espcrc1}
\usepackage{graphicx}

\def\eq#1{(\ref{#1})}
\def\Eq#1{(\ref{#1})}

\def\bea{\arraycolsep .1em \begin{eqnarray}}
\def\eea{\end{eqnarray}}

\def\s0#1#2{\mbox{\small{$ \frac{#1}{#2} $}}}
\def\0#1#2{\frac{#1}{#2}}
\def\beq{\begin{equation}}
\def\eeq{\end{equation}}

\title{Transport theory and low energy properties of colour
superconductors}

\author{
Daniel F.~Litim\\
Theory Group, CERN, CH -- 1211 Geneva 23, Switzerland.\\[2ex]
{\footnotesize CERN-TH-2001-315 }}%

\begin{document}

\maketitle

\begin{abstract}
  The one-loop polarisation tensor and the propagation of
  ``in-medium'' photons of colour superconductors in the 2SC and CFL
  phase is discussed. For a study of thermal corrections to the low
  energy effective theory in the 2SC phase, a classical transport
  theory for fermionic quasiparticles is invoked.
\end{abstract}

\section{Introduction}
Quantum chromodynamics (QCD) under extreme conditions displays a very
rich phase structure. At low temperature and high baryonic density,
quarks form Cooper pairs due to attractive interactions amongst them.
The diquark condensate modifies the ground state of QCD and leads to
the phenomenon of colour superconductivity \cite{Rajagopal:2000wf}.
This phase is typically characterised by the Anderson-Higgs mechanism
and an energy gap associated to the fermionic quasiparticles. It is
expected that a colour superconducting state of matter is realized in
compact stars.

For two light quark flavours the diquark condensate breaks the
$SU_c(3)$ group down to $SU_c(2)$ (2SC phase) \cite{Rajagopal:2000wf}.
Five gluons acquire a mass, while three gluons remain massless and
exhibit confinement.  Furthermore, quarks of one colour remain
ungapped, as they do not participate in the condensation. A global
axial $U_A(1)$ is broken at asymptotically large densities, leading to
a pseudo Nambu-Goldstone mode, similar to the $\eta'$ meson.  For
three light quark flavours the condensates lock the colour and flavour
symmetry (CFL phase) \cite{Rajagopal:2000wf}.  Colour, chiral and
baryon number symmetry are spontaneously broken.  All gluons become
massive and all quarks are gapped. The light degrees of freedom are
the Nambu-Goldstone bosons -- eight associated to the breaking of
chiral symmetry, one associated to the breaking of baryon number
symmetry, and, at asymptotically large densities, an extra pseudo
Nambu-Goldstone boson associated to the breaking of $U_A(1)$ -- which
dominate the long distance physics of the superconductor.

In this contribution, we discuss results obtained in collaboration
with C.~Manuel in \cite{Litim:2001mv,Litim:2001je}.
First, we include electromagnetic interactions and study the one-loop
photon polarisation tensor in colour superconductors
\cite{Litim:2001mv}.  Second, we discuss a semi-classical transport
theory for quark quasiparticles in the 2SC phase \cite{Litim:2001je}.
This is a convenient starting point for both the construction of low
energy effective theories or the computation of transport coefficients
of dense quark matter.

\section{Photons in colour superconductors}

A diquark condensate breaks, apart from the colour and flavour
symmetry, also the $U(1)$ symmetry of electromagnetism
\cite{Rajagopal:2000wf}. However, the linear combination ${\widetilde
  A}$ of the standard photon and the eighth gluon remains massless,
\bea
{\widetilde A}  =
- \sin{\theta}\, G^8 + \cos{\theta}\, A \,,\qquad
{\widetilde G}^8 =
 \cos{\theta}\, G^8+ \sin{\theta}\, A \,,\qquad
\widetilde e=e\cos{\theta}\,.
\eea
The field ${\widetilde A}$ plays the role of the ``in-medium'' photon
in the superconductor, with the in-medium gauge coupling $\widetilde
e$. Hence, the usual $U(1)$ charge generator $Q$ is replaced by the
in-medium generator $\widetilde Q$ \cite{Rajagopal:2000wf}. Notice
also that the rotation angle $\theta$ differs for the 2SC and CFL
phases,
\beq
\cos{\theta_{\rm CFL}} =  {g}\,(g^2 + \s043 e^2)^{-1/2} \,,
\qquad
\cos{\theta_{\rm 2SC}} =  { g}\,(g^2 + \s013 e^2)^{-1/2} \,.
\eeq
In the CFL phase, the charged particles are the Goldstone bosons
$\pi^{\pm}, K^{\pm}$, four gluons, and four quarks, all with
$\widetilde Q$-charge $\pm\widetilde e$. The condensates are
$\widetilde Q$-neutral. The spectrum contains the same number of
positively and negatively charged particles. In consequence, quark
matter in the CFL phase is $\widetilde Q$-neutral.  This holds also in
the presence of $K^0$ condensates. CFL matter is also $Q$-neutral,
even for a non-vanishing strange quark mass or chemical potential
\cite{Rajagopal:2001ff}. No further charge carriers are needed to make
the system electrically neutral.  In the 2SC phase, four gluons and
four quarks have half integer charges $\pm\s012\widetilde e$, while
the up quark which is not participating in the condensate has the
$\widetilde Q$-charge $\widetilde e$.  Additional heavy strange quarks
have the $\widetilde Q$-charge $-\s012\widetilde e$.  Again, the
condensates are electrically neutral. $\widetilde Q$-neutrality of the
full system requires additional negative charges, either as a
background of heavy strange quarks or electrons \cite{Rajagopal:2000wf}.

The photon self energy, to one-loop order, is given by all one-loop
diagrams with two external photon lines and internal loops of charged
particles.  The finite parts of heavy gluon loops are suppressed in
the infrared, as are those from the light pions and kaons.  Hence,
photon polarisation effects are entirely dominated in the infrared
limit by the quark loops \cite{Litim:2001mv}.
A simple physical picture has emerged from the explicit computation of
the photon polarisation tensor \cite{Litim:2001mv}. For intermediate
photon momenta $\Delta\ll p_0,p\ll\mu$, the polarisation tensor due to
gapped quarks reduces to its well-known hard-dense-loop counterpart in
the normal phase. This is easily understood. The photon wavelength is
larger than the size of the Cooper pair. Hence, the photons scatter
off the gapped quarks just as they would do in the phase without
condensation. They are Landau damped, static magnetic fields remain
unscreened while static electric fields are effectively Debye screened
with a mass of the order $\sim \widetilde e\mu$, due to all charged
quarks. In the 2SC phase, an additional contribution to the Debye mass
arises for non-vanishing electron chemical potential. For photons in
this momentum regime, and apart from numerical differences for the
value of the Debye mass and the couplings, the 2SC and CFL phases look
qualitatively the same.

The situation changes significantly for photons with low momenta
$p_0,p\ll\Delta$.  Here, their Compton wave length is too large to
resolve the condensates. The condensates are electrically neutral, but
they have an electrical dipole moment and hence modify the dielectric
constant $\widetilde \epsilon$ of the medium.  Condensates with
vanishing spin and angular momentum ($J=0$) have no magnetic moment.
Therefore, the magnetic permeability remains unchanged.  To leading
order in $p_0,p\ll\Delta$, one finds for the longitudinal and
transversal components of the polarisation tensor $\Pi^{\rm gap}$
arising from the gapped quarks \cite{Litim:2001mv}
\beq\label{gap}
\Pi^{\rm gap}_L (p_0, {\bf p})   =   - \widetilde\kappa\ p^2  \,, \qquad
\Pi^{\rm gap}_T (p_0, {\bf p})   =   - \widetilde\kappa\ p^2_0\,, \qquad
\widetilde\kappa                 =
\frac{ {c\,}}{18 \pi^2}
\frac{\widetilde e^2\mu^2}{\Delta^2} \,.
\eeq
The coefficient $c$ counts the charges squared of all gapped quarks
participating in the condensation: $c_{\rm 2SC}=1$ and $c_{\rm CFL}
=4$. The dielectric constant of the medium is
\beq\label{eps} \widetilde\epsilon = 1+\frac{ {c\,}}{18 \pi^2}
\frac{\widetilde e^2\mu^2}{\Delta^2} > 1\ .  \eeq
For realistic values of the gauge coupling, the gap and the quark
chemical potential, $\widetilde\kappa$ is of order one. It becomes
very large for asymptotic large densities and fixed $\widetilde e$.

In the CFL phase, the low momentum limit of the polarisation tensor is
solely given by \eq{gap}. Due to the absence of ungapped quarks,
electric fields are no longer Debye screened in the infrared limit.
Since all $\widetilde Q$-charged hadronic excitations in the CFL phase
acquire a gap for non-vanishing quark masses, the photon cannot
scatter if its energy is below the energy of the lightest charged
mode: CFL matter is a transparent insulator \cite{Rajagopal:2000wf}.
The polarisation of the medium implies that photons propagate with a
velocity smaller than the velocity of light in vacuum. The refraction
index of CFL matter with the normal phase is $\widetilde n=\widetilde
\epsilon^{1/2}>1$.

In the 2SC phase, the low momentum limit of the polarisation tensor is
given by \eq{gap} plus an additional contribution from the charged
ungapped quark and, possibly, electrons. In the static limit, they
combine into $\Pi_L=-(\widetilde m^2+ \widetilde\kappa\ p^2)$ with a
Debye mass due to the ungapped quark and the electrons.  Therefore,
photons still scatter from the charged fermions, and Debye screening,
though now with a different Debye mass, persists even for low momenta:
2SC matter is an opaque conductor.

To conclude, we have seen that the propagation of photons in a colour
superconducting phase is very different from the normal phase, and
also different for the different colour superconducting phases. This
result is an important input for a quantitative study of transport
coefficients like the electrical or thermal conductivity of the
medium. Although our results are based only on a one-loop computation,
the corrections to the dielectric constant are small for realistic
values of the parameters. Therefore, one may expect that higher loop
orders do not change the qualitative picture discussed here.

\section{Transport in colour superconductors}

Next, we study thermal corrections to the low energy physics of a
two-flavour colour superconductor \cite{Litim:2001je}. For $T=0$, an
infrared effective theory has been given recently for the massless
$SU(2)_c$ gauge fields \cite{Rischke:2000cn}. The gauge field dynamics
differs from the vacuum theory, because the condensates, although
neutral with respect to the unbroken $SU(2)$, polarise the medium
\cite{Rischke:2000ra}. This is fully analogous to what has been found
above for the photon interactions. For momenta $q\ll\Delta$, the
effective theory is
\begin{equation}\label{RSS}
S_{\rm eff}^{T=0}[A] = \int d^4 x \left ( \frac{\epsilon}{2} \,
{\bf E}_a \cdot {\bf E}_a - \frac{1}{2} \,{\bf B}_a \cdot {\bf B}_a
\right) \ ,
\label{Seff-T0}
\end{equation}
where $E_i^a \equiv F_{0i}^a$ and $B_i^a \equiv \frac12 \epsilon_{ijk}
F_{jk}^a$ are the $SU(2)$ electric and magnetic fields. Here,
$\epsilon = 1+ g^2 \mu^2/(18 \pi^2 \Delta^2)$ is the colour dielectric
susceptibility \cite{Rischke:2000cn}. The colour magnetic
susceptibility remains unchanged. This theory is confining with a
highly reduced scale of confinement $\Lambda'_{\rm QCD} \sim \Delta
\exp{(-\frac{2 \sqrt{2} \pi}{11} \frac{\mu}{g \Delta})}$. Due to
asymptotic freedom, it is expected that perturbative computations are
reliable for energy scales larger than $\Lambda'_{\rm QCD}$.

At non-vanishing temperature, thermal excitations modify the low
energy physics.  The condensate melts at the critical temperature $T_c
\approx 0.57 \Delta_0$, where $\Delta_0$ is the gap at vanishing
temperature. We restrict the discussion to temperatures within
$\Lambda'_{\rm QCD} \ll T <T_c$. In this regime, the main contribution
stems from the thermal excitations of the gapped quarks. Those of the
massless gauge fields are of the order $g^2T^2$ and subleading for
sufficiently large $\mu$, and the gapless quarks and the $\eta$ meson
do not couple to the $SU(2)$ gauge fields. We introduce a transport
equation for the quark quasiparticles, which are described in terms of
an on-shell one-particle phase space density $f(x,{\bf p},Q)$,
$x^\mu=(t,{\bf x})$. The distribution function depends on time, the
phase space variables position ${\bf x}$, momentum ${\bf p}$, and on
$SU(2)$ colour charges $Q_a$, with the colour index $a =1,2$ and $3$.
The quasiparticles carry $SU(2)$ colour charges simply because the
constituents of the condensate do.  Using natural units $k_B = \hbar =
c=1$, the on-shell condition for massless quarks $m_q=0$ relates the
energy of the quasiparticle excitation to the chemical potential and
the gap as $p_0 \equiv \epsilon_p$, with
\beq\label{velocity}
 \epsilon_p= \sqrt{(p-\mu)^2 + \Delta^2(T)}\,,\qquad
{\bf v}_p
\equiv \frac{ \partial \epsilon_p}{\partial {\bf p}} =
\frac{|p-\mu|}{\sqrt{(p-\mu)^2 + \Delta^2(T)}}\, {\hat {\bf p} \,.}
\eeq
Hence, the quasiparticle velocity ${\bf v}_p$ depends on both the
chemical potential and the gap. In the presence of the gap, their
propagation is suppressed, $|{\bf v}_p|< 1$.  The one-particle
distribution function obeys the transport equation
\begin{equation}\label{transport}
\left[D_t + {\bf v}_p \cdot {\bf D} -g Q_a \left({\bf E}^a +
{\bf v}_p \times {\bf B}^a \right) \frac{\partial}{\partial {\bf p}}
    \right] f = C[f] \,.
\end{equation}
Here, we have introduced the short-hand notation $D_\mu f\equiv
[\partial _\mu-g \epsilon^{abc}Q_c A_b^\mu{\partial ^Q_a}]f$ for the
covariant derivative acting on $f$. The first two terms on the
left-hand side of \Eq{transport} combine to a covariant drift term
$v_p^\mu D_\mu$, where $v_p^\mu = (1, {\bf v}_p)$ and $D_\mu =(D_t,
{\bf D})$. The terms proportional to the colour electric and magnetic
fields provide a force term. The right-hand side of \Eq{transport}
contains an unspecified collision term $C[f]$. Eq.~\Eq{transport} is
very similar to transport equations used earlier for hot or dense QCD
in the normal phase \cite{Heinz:1983nx,Litim:2001db}. The main
difference is the non-trivial dispersion relation \eq{velocity}, which
leads to modified Wong equations in the present case.  The thermal
quasiparticles carry a $SU(2)$ charge, and provide a $SU(2)$ colour
current. We introduce the colour density
\begin{equation}\label{currentdensity}
J_{a}(x,{\bf p})= g  \!\int\! dQ Q_{a} f(x,{\bf p},Q) \ ,
\end{equation}
where an implicit sum over species or helicity indices on $f$ is
understood.  The colour measure obeys $\int dQQ_a=0$ and $\int
dQQ_aQ_b=C_2\delta_{ab}$ ($C_2=\s012$ for quarks in the fundamental)
\cite{Litim:2001db}. The induced colour current of the medium follows
from \eq{currentdensity} as $J^\mu_a(x) = \int
\s0{d^3p}{(2\pi)^3}v_p^\mu J_{a}(x,{\bf p})$.  It is covariantly
conserved.

As an application of \eq{transport}, we study the collisionless
dynamics $C[f] =0$ close to thermal equilibrium and to leading order
in the gauge coupling.  Consider the distribution function $f= f^{\rm
  eq}+ g f^{(1)}$ where $f^{\rm eq} = (\exp\epsilon_p/T
+1)^{-1}$ is the fermionic equilibrium distribution function and $g
f^{(1)}$ a slight deviation from equilibrium. Expanding the transport
equation \eq{transport} in $g$, and taking the two helicities per
quasiparticle into account, we find
\begin{equation}\label{transport-current}
\left[D_t + {\bf v}_p \cdot {\bf D}\right]   J(x,{\bf p})  =
g^2 N_f  \,  {\bf v}_p \cdot  {\bf E}(x)\, \frac{d f^{\rm eq}}{d
\epsilon_p} \ .
\end{equation}
This equation can be solved for $J(x,{\bf p})$ and gives the full
colour current as a functional of the gauge fields only,
$J(x)=J[A](x)$ \cite{Litim:2001je}. It defines the ''hard
superconducting loop'' effective action $\Gamma_{\rm HSL}[A]$ through
$J[A] = -{\delta \Gamma_{\rm HSL}[A]}/{\delta A}$, in full analogy to
the derivation of the hard thermal loop effective theory within
classical transport theory \cite{Litim:2001db}.  Hence, the low energy
effective theory for modes with $q\ll \Delta$ in the 2SC phase at
finite temperature is $S_{\rm eff}^T[A] = S_{\rm eff}^{T=0}[A] +
\Gamma_{\rm HSL}[A]$ to leading order in $g$.

Consider the thermal polarisation tensor which
follows from $J[A]$ \cite{Litim:2001je}. It describes Landau
damping and Debye screening of the $SU(2)$ gauge fields in the 2SC
phase. For low momenta, it agrees with the findings of
\cite{Rischke:2000ra}, based on a field-theoretical derivation. In
contrast to the normal phase, the dispersion
relations for longitudinal and transverse gluons are modified. For
example, the Debye mass is $m^2_D=(2\Delta/\pi^2T)^{1/2} g^2\mu^2 \exp
-\Delta/T$ in the limit of large density and low temperature.  In the
general case, the Debye mass can only be computed numerically.  An
approximate expression for the polarisation tensor could be given in
terms of an effective mean squared velocity $v_*$ for the
quasiparticles, similar to the approximation used in
\cite{Braaten:1993jw} for the normal phase.

In conclusion, we have discussed a very simple classical transport
equation for quark quasiparticles in the 2SC phase. To leading order
in $g$, it leads to the same result as a full field theoretical
computation.  Due to its simplicity, the present approach seems to be
a good starting point for the computation of transport coefficients in
the 2SC phase, like the colour conductivity. Here, the techniques of
classical transport theory can be exploited
\cite{Heinz:1983nx,Litim:2001db,Litim:1999ns}. It would also be
interesting to provide a similar approach for the CFL phase, where the
low energy degrees of freedom are substantially different.
\\[-.5ex]

{\it Acknowledgements:} I thank C.~Manuel for collaboration, and the
organisers for a stimulating conference and financial support. This
work has been supported by the European Community through the
Marie-Curie fellowship HPMF-CT-1999-00404.


\begin{thebibliography}{99}


\bibitem{Rajagopal:2000wf}
See K.~Rajagopal and F.~Wilczek,
hep-ph/0011333 for a review and relevant literature.

\bibitem{Litim:2001mv}
D.~F.~Litim and C.~Manuel,
Phys.\ Rev.\ D {\bf 64} (2001) 094013
[hep-ph/0105165].

\bibitem{Litim:2001je}
D.~F.~Litim and C.~Manuel,
Phys.\ Rev.\ Lett.\  {\bf 87} (2001) 052002
[hep-ph/0103092].


\bibitem{Rajagopal:2001ff}
K.~Rajagopal and F.~Wilczek,
Phys.\ Rev.\ Lett.\  {\bf 86} (2001) 3492
[hep-ph/0012039].


\bibitem{Rischke:2000cn}
D.~H.~Rischke, D.~T.~Son and M.~A.~Stephanov,
Phys.\ Rev.\ Lett.\  {\bf 87} (2001) 062001.

%
\bibitem{Rischke:2000ra}
D.~H.~Rischke,
Phys.\ Rev.\  {\bf D62} (2000) 054017
[nucl-th/0003063].

\bibitem{Heinz:1983nx}
U.~W.~Heinz,
Phys.\ Rev.\ Lett.\  {\bf 51} (1983) 351;
Annals Phys.\  {\bf 161} (1985) 48.

\bibitem{Litim:2001db}
D.~F.~Litim and C.~Manuel,
hep-ph/0110104.

\bibitem{Braaten:1993jw} E.~Braaten and D.~Segel,
  Phys.\ Rev.\ D {\bf 48} (1993) 1478 [hep-ph/9302213].

\bibitem{Litim:1999ns}
D.~F.~Litim and C.~Manuel, Phys.\ Rev.\ Lett.\  {\bf 82} (1999) 4981
[hep-ph/9902430];
Nucl.\ Phys.\ B {\bf 562} (1999) 237
[hep-ph/9906210].

\end{thebibliography}
\end{document}